\documentclass[10pt, conference, compsocconf]{IEEEtran}

\usepackage{amssymb}
\usepackage{algorithm}
\usepackage[bookmarks=false]{hyperref}
\usepackage{amsfonts, amsmath, amsthm, amssymb}
\usepackage{float}
\usepackage{subcaption}
\usepackage{makecell}
\usepackage{booktabs}
\usepackage{paralist}
\usepackage{xcolor}
\usepackage{booktabs}

\ifCLASSINFOpdf
 \usepackage[pdftex]{graphicx}
 \graphicspath{{images/}}
\else
\fi
\graphicspath{ {images/} }
\DeclareGraphicsExtensions{.png}

\newcommand\todo[1]{{\textbf{[TODO: #1]}}}

\newcommand{\boldpar}[1] {\vspace{0.3cm}\noindent\textbf{#1}~~}

\begin{document}


\title{ \LARGE \textbf{
   In Datacenter Performance, The Only Constant Is Change
}}

\author{
  \IEEEauthorblockN{
  Dmitry Duplyakin\textsuperscript{*},
  Alexandru Uta\textsuperscript{$\dagger$}\textsuperscript{$\diamond$},
  Aleksander Maricq\textsuperscript{*},
  Robert Ricci\textsuperscript{*}
  }
  \IEEEauthorblockA{
  \textit{\normalsize{\textsuperscript{*}University of Utah,
  \textsuperscript{$\dagger$}Vrije Universiteit Amsterdam, \textsuperscript{$\diamond$}Leiden University}}
  \\
  \textsuperscript{*}\{dmdu, amaricq, ricci\}@cs.utah.edu,
  \textsuperscript{$\dagger$}a.uta@vu.nl
  }
}

\maketitle

\IEEEpeerreviewmaketitle

\begin{abstract}

All computing infrastructure suffers from performance variability, be it bare-metal or virtualized. This phenomenon originates from many sources: some transient, such as noisy neighbors, and others more permanent but sudden, such as changes or wear in hardware, changes in the underlying hypervisor stack, or even undocumented interactions between the policies of the computing resource provider and the active workloads.
Thus, performance measurements obtained on clouds, HPC facilities,
and, more generally, datacenter environments are almost guaranteed to
exhibit performance regimes that evolve over time, 
which leads to undesirable nonstationarities in application performance.
In this paper, we present our analysis of performance of the bare-metal hardware 
available on the CloudLab testbed where we focus on quantifying the evolving performance regimes using changepoint detection. 
We describe our findings, backed by a dataset with nearly 6.9M benchmark results collected from over 1600 machines over a period of 2 years and 9 months.
These findings yield a comprehensive characterization of
real-world performance variability patterns in one computing facility, 
a methodology for studying such patterns on other infrastructures, 
and contribute to a better understanding of performance variability in general.

\end{abstract}


\section{Introduction}

The performance of computing infrastructure is \emph{variable},
which means multiple runs of the same code on the same hardware result in slightly different performance results~\cite{Maricq+:OSDI18, jain}. Often, this variation is relatively small (within a few percent) and follows a pattern that can be modeled as \emph{random noise}, with samples following a \emph{stationary} (though potentially complicated) distribution. Under these conditions, it is possible to use sound experiment designs~\cite{santner2003design, balaprakash2013active, duplyakin2016active} and statistical techniques~\cite{jain, hoefler2015scientific} to achieve meaningful performance results.

Sometimes, however, the performance distribution does \emph{not} remain stationary: it exhibits some systematic change such as the change in the median, variance, tail, or other statistical properties. Instances of such change are called \emph{changepoints}, and the practice of finding them is referred to as \emph{Changepoint Detection} (CPD)~\cite{page1954continuous,smith1975bayesian}. 
In this study, we consider the problem of detecting performance changepoints in large computing facilities. We use data collected from CloudLab, 
a distributed testbed which supports
cloud computing and systems research by providing raw access to programmable
hardware~\cite{Duplyakin+:ATC19}.
Using 6.9M measurements collected on more than 1600 bare-metal machines
during a period of 2 years and 9 months 
(May 2017--February 2020), 
we detect between 583 and 2439 changepoints (depending on the detection sensitivity).
In this paper, we describe our assessment of the magnitude of these performance changes and the duration of inter-changepoint intervals, as
well as relate groups of changepoints to 
several major recorded configuration changes that occurred on CloudLab
over the course of this benchmarking effort.

Understanding changepoints in facility performance is helpful to both operators and users. In operation of a large-scale facility, change over time is a fact of life: our analysis in Section~\ref{cpd} shows that performance change is the norm rather than the exception. Hardware ages, firmware and software get updated, security patches~\cite{databricks-spectre} are applied, and even changes in the physical environment affect performance over time~\cite{gunawi18failslow}. Some of these changes are planned, some are not, and some changes have unexpected performance consequences. The level of ``noise'' in repeated performance measurements and between different machines in the same facility makes detecting and characterizing these changes difficult. However, our work should help operators better understand their own facilities by giving them insights into the effects of planned system updates, allowing them to find unexpected changes,
and helping them grasp which metrics change together.

On the user side, changepoints can help users better understand performance of their own codes and run repeatable experiments. When one gets unexpected performance results, a natural question to ask is ``Has something on the platform changed?'' Our CPD-based approach helps answer this type of questions: given some performance measurements collected between two points in time, we can automatically detect significant changepoints that occurred between them and identify the sets of impacted metrics, as well as the magnitudes of the observed changes. This is also important for \emph{repeatable} and predictable performance research~\cite{patki2019performance}: if the goal is to compare the performance of two programs where one of them was originally run months or years ago, it is essential to understand if the baseline performance of the platform has evolved since that time, and, if so, in what specific ways and by how much.

In the face of non-trivial and often unexpected performance changes, practitioners are often unprepared, especially when they do not control the underlying infrastructure. We advocate for constantly \emph{fingerprinting} the underlying resources' performance as a prerequisite to \emph{application performance explainability}. Similar to the recent concept of developing explainable artificial intelligence algorithms, making sense of systems' performance is becoming an extremely complex yet necessary endeavor, especially at large scale. Therefore, performance fingerprinting, which includes frequent benchmarking of
individual components of computing infrastructure, 
is imperative to the explicable system performance behavior.

As shown in recent studies, performance evaluations in the literature do not always use experimental practices that account for variability and change over time~\cite{uta2020NSDI}. Building benchmarks that automatically adjust based on performance variability leads to more trustworthy evaluations~\cite{kogias2019lancet} and understanding of the types and frequency of changes that occur in practice is vital to this adaptation. Additionally, understanding changes over time---especially unplanned ones---can help system designers build systems that are stable and offer consistent performance guarantees.

Toward persuading the international systems community that in datacenter performance, stability is usually the exception rather than the norm~\cite{Maricq+:OSDI18}, and that better experimental design and practices are needed~\cite{uta2020NSDI,uta2018performance}, in this paper we make the following contributions:
\begin{enumerate}
    \item We advocate for and describe the power of CPD in characterizing performance variability (Section~\ref{cpd}).
    \item We characterize and identify changepoints in the performance of CloudLab. We offer clear-cut examples for validating our CPD results, such as OS and BIOS patches (Section~\ref{investigating}).
    \item We offer practitioners and performance engineers guidelines toward tuning the sensitivity of CPD and setting expectations 
    about its performance (Section~\ref{tuning}).
    \item For promoting reproducibility, we release as open data a large-scale archive containing several million datapoints characterizing the CPU, memory, and disk performance of 
    a heterogeneous pool of hardware resources available on CloudLab, 
    as well as our analysis tools (Section~\ref{artifacts}).
\end{enumerate}

\noindent To give the aforementioned contributions more context,
we describe the dataset we have collected in Section~\ref{benchmarking}
and discuss CPD-related developments with potential high impact in
computer systems research in Section~\ref{impact}.

\section{Related Work / Background}
\label{related}

Identifying changepoints in datasets is regarded as a useful technique to pinpointing when certain characteristics of the recorded data have changed. More precisely, a changepoint is a temporal moment that separates a given dataset in two sub-datasets with different statistical characteristics. Such analysis has proven itself valuable in several types of analyses related to large-scale systems. In this section, we start by giving a brief overview of scalable changepoint detection techniques and continue by describing the applicability domains of CPD.

\textbf{Changepoint Detection Techniques.} There are frequentist~\cite{page1954continuous,page1955test} and Bayesian~\cite{smith1975bayesian,adams2007bayesian,stephens1994bayesian} approaches to changepoint detection, and it is shown that both of them achieve good results in online and offline settings. Although computationally demanding, there exist scalable algorithms and efficient software packages for CPD~\cite{scott1974cluster,killick2012optimal,killick2014changepoint}. Such linear or sub-quadratic time algorithms ensure that even massive datasets can be analyzed quickly, in a scalable way and in near real-time scenarios.

\textbf{Changepoint Detection in Large-scale Systems.} In large-scale systems, changepoint detection has been used in a variety of scenarios, including decentralized sensor networks~\cite{tartakovsky2003quickest}, performance diagnosis in distributed systems~\cite{chen2014causeinfer}, identifying JVM warmups~\cite{barrett2017virtual}, denial-of-service attacks and intrusion detection~\cite{chen2007distributed,blazek2001novel,tartakovsky2006novel}. These techniques prove scalable and efficient enough to inform real-time decisions in time-critical applications.

\textbf{Anomaly Detection in Large-scale Systems.} Computer systems are generally designed with clear (non-)functional requirements. Engineers and administrators measure such metrics to determine when systems deviate from their normal behavior, i.e., detecting anomalous behavior. Various changepoint detection methods~\cite{li2010fast,wang2011statistical,solaimani2014spark,tan2011fast} are also successfully used for detecting anomalies. 

Highly contrasting with our work, all the previously discussed research considers changepoints or anomalies the exception rather than the norm. However, in this paper we make the case for \emph{change being the norm} in datacenter performance. Toward understanding this behavior, we show strong empirical evidence and we advocate for continuously monitoring for performance changes. By accounting for change, engineers, datacenter operators, and researchers can design better benchmarks, easily explain abnormal or seemingly inexplicable performance behavior and build better datacenters and distributed applications.

\section{Large-Scale Benchmarking}
\label{benchmarking}

\subsection{Previous Work}
Our initial work in the area of facility-wide hardware performance analysis started in May 2017 with a study of
hardware available on the CloudLab testbed.  CloudLab allocates an entire bare-metal 
machine to one user at a time, and therefore we can study the variability of the hardware performance free from the side-effects of 
``noisy neighbors'' and virtualization~\cite{novakovic2013deepdive}.  
More specifically, we were interested in (i) \emph{how the performance differs between 
supposedly identical machines} and (ii) \emph{how the same machine performs over time}.  
Over a period of 10 months, we captured nearly 900K data points from 835 machines by 
testing subsets of available machines several times per day. Machines were tested using network, memory, and disk microbenchmarks suites according to the least-recently-tested order.  
Our observations and statistical analysis resulted
in a publication where we described thirteen data-backed findings from this study, spanning both experimentalists' and facility operators' concerns~\cite{Maricq+:OSDI18}.

Since our initial collection period, we expanded the scope of this performance analysis to 
include the newest hardware and additional benchmarks, as well as increased the number of machines tested per collection period.  
As of the writing of this paper, the dataset has grown to include over 
4.3M memory and 664K disk performance measurements from over 1600 machines.
We have also gathered over 2.0M CPU test results
and reported initial findings from our high-level
assessment of the observed variability patterns~\cite{Duplyakin+:MERIT19}.

\subsection{Ongoing Benchmarking and Analysis}

Close to our current work in terms of experiment scale are the studies of Gunawi et al.~\cite{gunawi18failslow} and Amvrosiadis et al.~\cite{amvrosiadis2018diversity}. The former investigates the symptoms of ``fail-slow'' hardware and characterize the performance anomalies generated by such issues. The latter shows that trace-driven experiment designs can lead to over-fitting of systems software toward certain behavior in case not enough (varied) traces are used. Much like those studies, we investigate a rather unexplored path in our domain, namely 
performance analysis of cyberinfrastructure with the focus on 
CPD.
This analysis uncovers non-trivial behavior in performance measurements and helps characterize change patterns in datacenter performance.
We present the data and the findings that have potential 
    to inform new research in the area of \emph{adaptive benchmarking},
    represented by studies such as \cite{kalibera2013rigorous} and \cite{kogias2019lancet}, and should promote the development of CPD-based tools,
    both for analysis and systems management.

This study expands our previous temporal analysis of performance data~\cite{Maricq+:OSDI18} in two important ways.
First, we go beyond simply checking stationarity with the Augmented Dickey--Fuller test~\cite{dickey1979distribution},
    which previously showed that most of our performance traces cannot be 
    viewed as stationary with high conference level.
We describe an investigation that involves quantifying individual changepoints and temporary steady states.
Second, we describe the analysis 
    that not only characterizes individual timeseries
    but also processes entire batches of performance timeseries
    and helps reveal relationships between 
    different performance manifestations of the same system events.
By relating and grouping different changepoints we identify, 
    we gain better understanding of the underlying causes. 

\subsection{Collected Performance Data}

\begin{table}[t]
    \centering
    \caption{Hardware specs and test coverage.}
    \resizebox{\columnwidth}{!}{
    \begin{tabular}{|l|l|l|l|l|l|l|}
        \hline
        \textbf{Type} & \textbf{Model} & \textbf{Processor} & \textbf{Cores} & \textbf{Tested / Total \#} & \makecell{\textbf{Measurements} \\ \textbf{CPU / Memory}} \\
        \hline
        \texttt{m400} & HPE m400 & APM X-GENE & 8 & 311 / 315 & 151 K / 201 K \\ \hline
        \texttt{m510} & HPE m510 & Xeon D-1548 & 8 & 266 / 270 & 214 K / 575 K \\ \hline
        \texttt{xl170} & HPE xl170r G9 & Xeon E5-2640v4 & 10 & 196 / 200 & 152 K / 270 K \\ \hline
        \texttt{c220g1} & Cisco c220m4 & Xeon E5-2630v3 & 16 & 89 / 90 & 121 K / 375 K \\ \hline
        \texttt{c220g5} & Cisco c220m5 & Xeon Silver 4114 & 20 & 220 / 224 & 380 K / 675 K \\ \hline
        \texttt{c6220} & Dell C6220 & Xeon E5-2650v2 & 16 & 60 / 64 & 98 K / 175 K\\ \hline
        \texttt{c6320} & Dell C6320 & Xeon E5-2683v3 & 28 & 84 / 84 & 156 K / 444 K\\ \hline
        \texttt{c6420} & Dell C6420 & Xeon Gold 6142 & 32 & 71 / 72 & 59 K / 105 K  \\ \hline
    \end{tabular}}
    \label{table:hw}
    \vspace*{-0.4cm}
\end{table}

We measure CPU performance of CloudLab hardware using NPB,
    the NAS Parallel Benchmarks~\cite{bailey1995parallel}, version 3.3--OMP.
We run 9 microbenchmarks (BT, CG, EP, FT, IS, LU, MG, SP, UA)
    on homogeneous pools of machines of 11 types,
    turning on/off dynamic voltage and frequency scaling (DVFS).
We vary the number of running threads---we run tests that use either a single thread or all available hardware threads---and also pin the computations to each of the sockets (for two-socket machines)
    using the \texttt{numactl} utility.
All these parameters create an input space with 590 distinct \emph{configurations} that we use to answer questions related to performance     changepoints in the current work.
Each run produces a record in our dataset
    with the runtime (in seconds)
    accompanied by many metadata attributes,
    which include machine specs, OS version, kernel release, and compiler version, among others.

Similarly, we collect measurements for 1038 memory configurations
    that correspond to evaluating the same CloudLab hardware using STREAM~\cite{McCalpin1995} tests
    and the micro-benchmarks from Alex W. Reece's suite~\cite{awreece-code,awreece-writeup} for
    testing Intel x86 intrinsics such as SSE and AVX instructions.
In Table~\ref{table:hw}, we detail some of the CloudLab's hardware types---the ones we refer to throughout the paper---their specifications, and the testing coverage.
The complete hardware overview can be found on
    the CloudLab Hardware documentation page~\cite{cloudlab-hw-docs}.

We include 152 disk configurations in our analysis. 
We run \texttt{fio}~\cite{fio} in different settings
    on raw I/O devices or their partitions,
    eight tests per device:
read and write load,
    random and sequential tests, 
    low and high $iodepth$ settings.
These configurations include the measurements
    for HDDs, SSDs, and NVMe devices,
    which provides opportunities for comparing
    entire classes of I/O devices.

We have made this entire 6.9M-measurement performance dataset publicly available as described in Section~\ref{artifacts}.

\section{Changepoint Detection Analysis}
\label{cpd}

\subsection{Overview of CPD}

\begin{figure*}[t]
    \captionsetup[subfigure]{justification=centering}
    \centering
    \begin{subfigure}{0.32\textwidth}
      \centering
      \includegraphics[clip, width=0.9\textwidth]{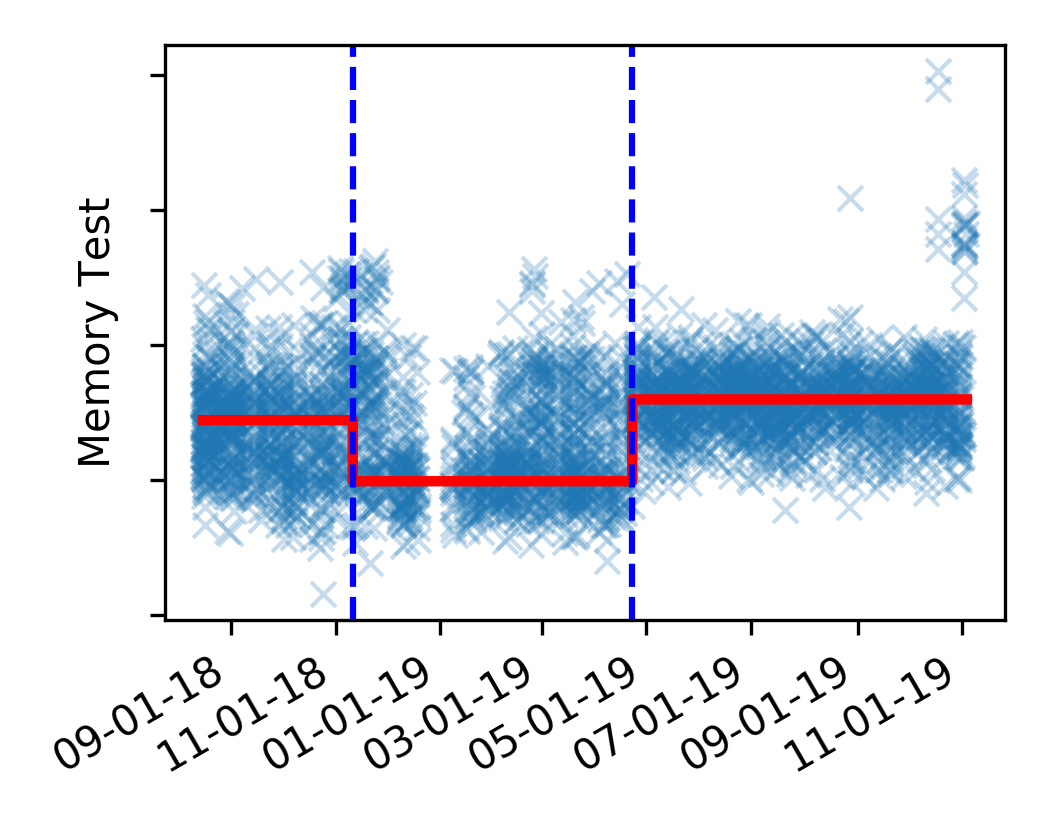}
      \caption{\small{Multi-threaded memory copy results.}}
      \label{fig:cpd-mem-all}
    \end{subfigure}%
    \begin{subfigure}{0.32\textwidth}
      \centering
      \includegraphics[clip, width=0.9\textwidth]{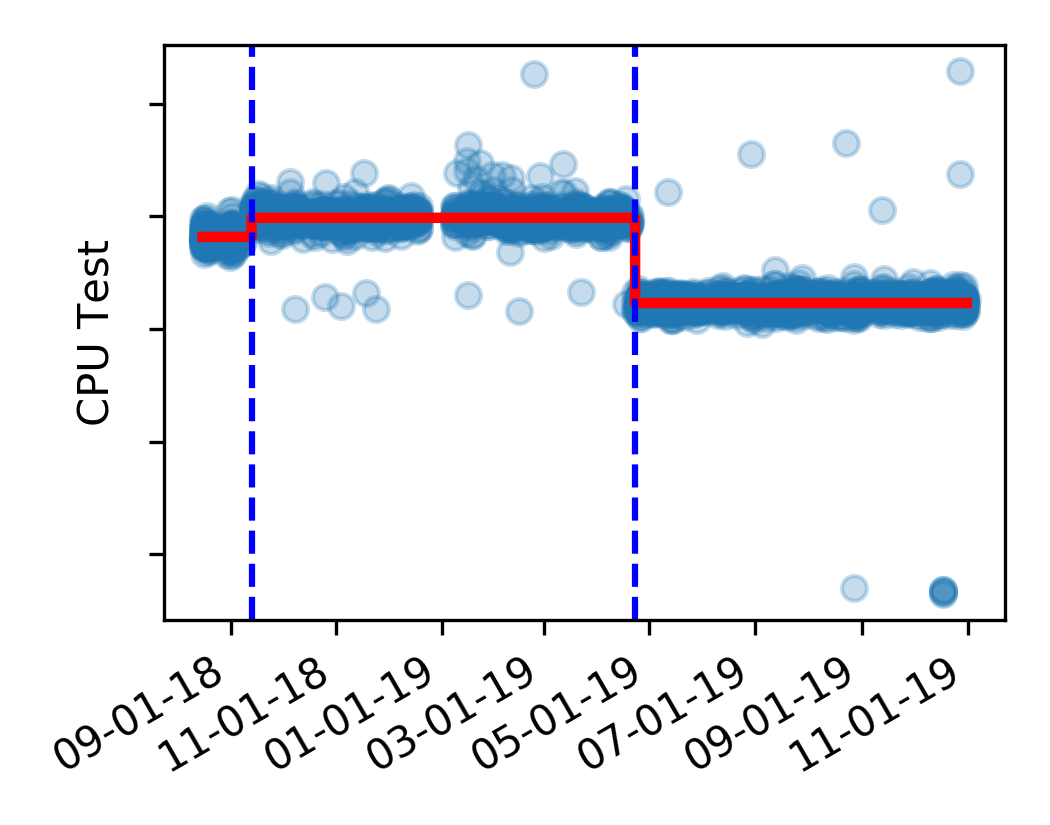}
      \caption{\small{Runtimes of multi-threaded FFT.}}
      \label{fig:cpd-cpu-all}
    \end{subfigure}
    \begin{subfigure}{0.32\textwidth}
      \centering
      \includegraphics[clip, width=0.9\textwidth]{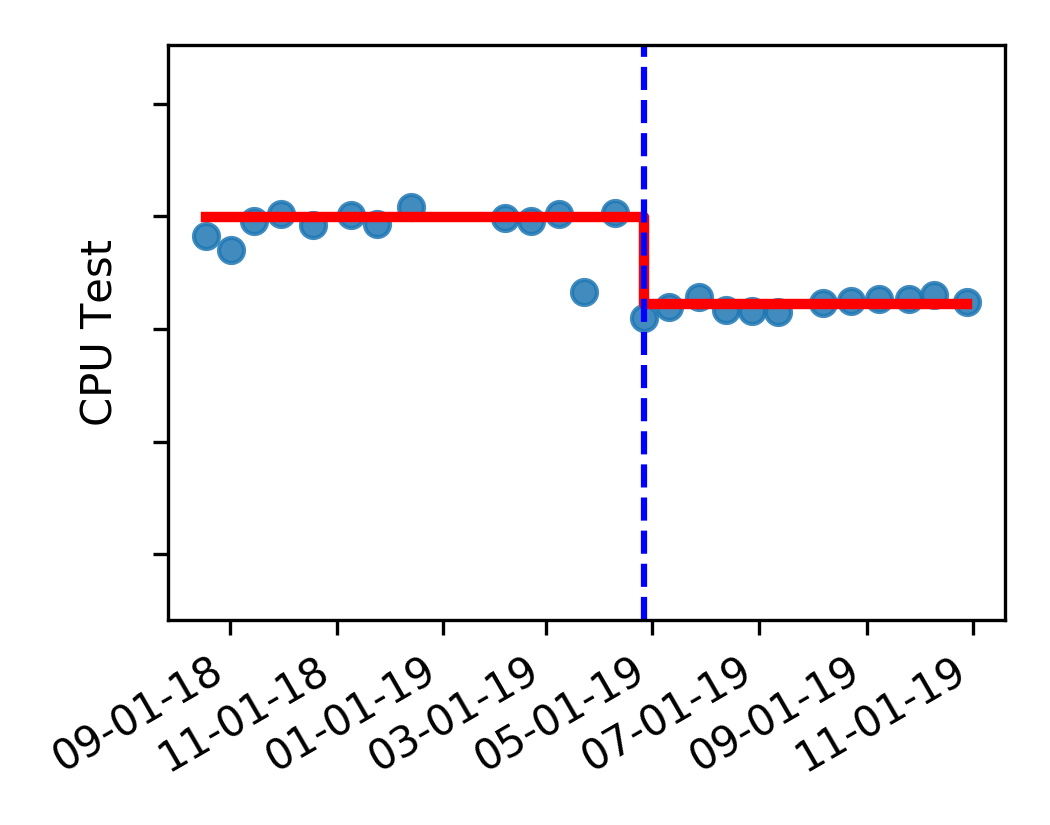}
      \caption{\small{Subset from (b) for a single machine.}}
      \label{fig:cpd-cpu-node}
    \end{subfigure}
    \caption{Outlier-resilient change-point detection applied to memory and CPU performance measurements
    collected from 183 homogeneous \texttt{xl170} machines
    over the period of 14 months, which started when these machines were first added to CloudLab.}
    \label{fig:cpd}
\end{figure*}

We leverage a recent CPD approach that is explicitly designed to handle data with outliers and heavy-tailed noise~\cite{fearnhead2019changepoint}.
This approach uses an efficient dynamic programming algorithm to produce
    minimum-cost \emph{segmentations} of univariate timeseries.
Not only is this approach shown to be robust to noise in the data,
    but it operates sequentially on the data that is given to it for processing,
    which makes it a great fit for rapid analysis of multiple streams of performance 
    measurements coming from large-scale cyberinfrastructure.
Robustness to noise is equally important in this context
    considering that outliers occur in performance results
    even on perfectly fine hardware due to nondeterminism 
    of computing systems and integral low-level attributes of their design, such as 
    task scheduling, interrupts, instruction pipelining, caching, among others
    \cite{Maricq+:OSDI18, hoefler2015scientific}.
Therefore, we base our work on the method 
    for which a handful of outliers, including large ones,
    would not result in changepoints being reported 
    unless they provide sufficient statistical basis for such outcomes.
We also explore how \emph{sensitive} the detection method we selected
    is to short-term fluctuations and report our finding throughout the paper.

We use an implementation of this CPD that is available in the form 
    of the \texttt{robseg} package in R~\cite{robust-fpop}. 
We integrate it with the analysis and visualization
    tools we develop in Python using the \texttt{rpy2} interface~\cite{gautier2008rpy2}.
Following one of the repository's examples,
    we run this implementation with the \emph{biweight} loss function 
    (i.e., pointwise minimum of an $L_2$ loss and a constant),
    which is shown to improve the consistency and the accuracy of changepoint estimation~\cite{fearnhead2019changepoint}.
The authors argue that this is a good choice in practice
    that has no performance drawbacks comparing to the
    more common and yet less outlier-resilient $L_2$ (square error) loss. 
They also introduce the penalty/threshold parameter $K$,
    which relates the magnitudes of potential changepoints 
    to the ratio of the signal to its standard deviation,
    and use different values with different loss functions. 
After our initial experiments with CloudLab performance traces,
    we have settled on the $[0.3, 1.0]$ range for $K$ values,
    making this \textit{hyperparameter} easily tunable in the analysis dashboard 
    we develop, as we discuss in Section~\ref{artifacts}. 
Most of the results we present in this paper are obtained with $K=0.6$, 
    unless noted otherwise. 
We would also like to note that while the authors of this CPD method
    evaluate it on the data with heavy-tailed but synthetic $t$-distributed noise
    and also empirical data from a well drilling application~\cite{ruanaidh2012numerical},
    we apply this recently developed CPD approach in the area of 
    computer performance analysis and, to the best of our knowledge, report
    on the first large-scale analysis of this kind.

The aforementioned choices allowed us to achieve desired results
    in segmentation of CloudLab's performance data
    in various settings, as demonstrated in Figure~\ref{fig:cpd}.
Both memory (a) and CPU (b,c) datasets were segmented 
    in the ways that made perfect sense visually and proved to be outlier-resilient.
The latter can be noticed in the memory plot (a)
    where some of the most recent measurements appeared higher than the rest,
    while the detection algorithm did not create another changepoint 
    that represents them; similar instances can be observed in the plot (b).
Either more data is needed to confirm the significance of this behavior,
    or we can increase the value of threshold $K$ if we 
    are indeed interested in capturing such instances.

Another interesting result here is that all three 
    of the shown segmentations---two for the data from all 
    machines of the studied hardware type (a,b) and one for a single machine of this type (c)---agree on the change that occurred just before ``05-01-19". 
The agreement between different benchmarks' changepoints speak 
    for the significance of such changes:
the higher the number of agreeing benchmarks,
    the more ``weight'' these performance changes carry. 
We search for examples of cluster-wide performance 
    changes using CPD and discuss several specific cases in more detail in Section~\ref{investigating}.
In this study, we focus on the analysis of
    traces that include large sets of measurements from batches of homogeneous machines,
    like traces (a) and (b).
We acknowledge that there is an alternative approach
    which involves processing smaller single-machine traces, similar to (c),
    one-by-one and then clustering the detected changepoints
    for extracting prominent patterns.
This method may provide its own benefits 
    (e.g., resource-specific CPD that may help expose outliers),
    yet it falls outside the scope of our current work.

We run the \texttt{robseg}-based CPD on the data for all CPU,
    memory, and disk configurations described in the previous section 
    and analyze the properties 
    of the produced segmentations, defined as follows.
The $n$th segment---the period of testing between changepoints
    where performance measurements appear stationary---can be characterized by the duration $d_n$ and the mean $m_n$ of the measurements that fall within it.
The first and the last segments in each time series 
    have the beginning or the end of the testing period as one of its end points.
Thus, the segmentations shown in Figure~\ref{fig:cpd}
    include the total of 8 segments (3 in (a), 3 in (b), and 2 in (c)).
At each changepoint, we observe a ``step'' that we can characterize as a relative change in the means:
$c_n = (m_{n+1}-m_{n}) / m_{n} \times 100\%$.
Below we describe the distributions of empirical $c_n$ and $d_n$ values we obtain for different benchmarks.

\subsection{Changepoints and Their Characteristics}

\begin{figure*}[t]
    \centering
      \includegraphics[clip, width=0.96\textwidth]{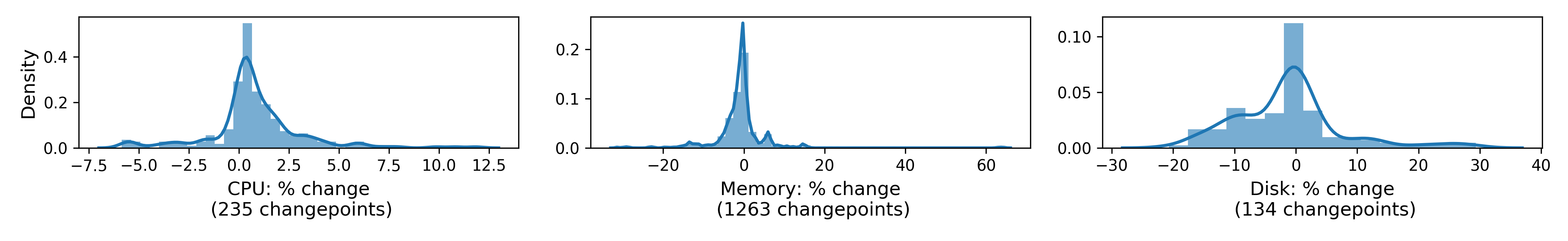}
    \caption{Histograms of $c_n$ values, the relative mean changes corresponding to the detected changepoints.}
    \label{fig:magnitudes}
\end{figure*}

\begin{figure*}[t]
    \centering
      \includegraphics[clip, width=0.96\textwidth]{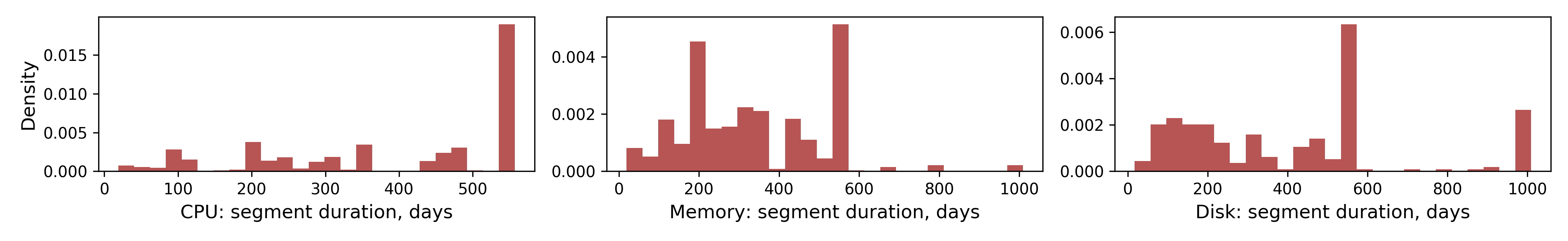}
    \caption{Histograms of $d_n$ values, the durations of segments with stable performance. The set of summarized changepoints is the same as the 
    set shown in Figure~\ref{fig:magnitudes}, which includes the total 
    of 1632 changepoints.}
    \label{fig:durations}
\end{figure*}

All changepoints we detect form three groups,
    representing CPU, memory, and disk performance changepoints.
The corresponding $c_n$ distributions that characterize 
    the relative magnitudes of these changes are shown in 
    three histograms in Figure~\ref{fig:magnitudes}.
Roughly speaking, most CPU changes are within the $[-7.5\%, 7.5\%]$ range,
    memory changes---within $[-20\%, 20\%]$,
    and disk changes---almost entirely within $[-30\%, 30\%]$.
We also notice that the directions of the heavier tails 
    in these distributions \emph{agree} with the directions
    of typical outliers in these types of performance measurements.
Thus, CPU changepoints have more positive $c_n$ values
    than negative
    (i.e., the median is slightly greater than zero),
    and the abnormal CPU performance results (measured with test runtimes, in units of time)
    appear on the high side of the usual performance levels.
In contrast, for memory and disk bandwidth tests (measured in MB/s),
    more $c_n$ values have the negative sign (and medians are below zero),
    which corresponds to instances of degradation of such performance metrics---the 
    scenario that is more common in practice for unexpected changes  
    than bandwidth improvements.
In Section~\ref{investigating}, we further discuss several specific 
    multi-benchmark changes that conform with these patterns 
    and also represent several counterexamples.
    
We also track the numbers of detected changepoints, 
    for each distribution and the total number.
They are directly proportional
    to the value of $K$, and the number of memory changepoints
    is greater than the number of CPU changepoints, which, in turn,
    is greater that the number of disk changepoints in our dataset.
Considering that we analyze uneven numbers of configurations     
    across these types, we adjust for their numbers 
    ($n^{CPU}=590$, $n^{Memory}=1038$, and $n^{Disk}=152$)
    and, using the notation $r^{X}=|\{c^{X}_n\}| / n^{X}$ for ratios of changepoints per configuration,
    arrive at the following:
    \[r^{Memory} > r^{Disk} > r^{CPU}.\]
This holds true for all values of $K$ in the range we have studied.
To provide a concrete example, 
    our analysis for $K=0.6$
    yields $r^{CPU}=0.40$, $r^{Memory}=1.22$, and $r^{Disk}=0.88$.
This suggests that on average in this setting,
    4 memory traces have approximately 5 changepoints,
    10 disk traces have 9 changepoints,
    and 5 CPU traces have 2 detectable changepoints.
In combination with the analysis of inter-changepoint intervals presented below,
    this fact should provide a reference point 
    for researchers and practitioners
    pursuing performance-focused changepoint detection,
    especially at large scale.

\subsection{Steady States}     

The inter-changepoint intervals with stationary performance regimes can be viewed 
    as \emph{steady states} (provided there is a representative set of measurements).
With these intervals, we are interested in the patterns expressed
    in the distributions of their durations.
If we find them to include many short intervals, we may consider tuning CPD
    to produce fewer changepoints and only characterize more permanent ones.
At the other extreme, with less sensitive CPD we run risk of not
    noticing some short-term but important changes.
In Figure~\ref{fig:durations}, we present what we believe are the \emph{balanced} 
    duration distributions, which we obtain for $K=0.6$ and measure in days.
Specifically, we analyze the heights of the leftmost bars in these histograms
    (i.e., characterizing short segments)
    and notice that they are comparable to the heights of 
    the bars representing much longer segments,
    at several hundred days.
In other words, the short-term changes are neither too abundant nor lacking,
    when compared to the changes that occur at larger timescales.  
With this summary in mind, we gain more confidence about  
    viewing the steady states on the shorter end
    of this spectrum
    as being representative of the system's performance regimes rather than stemming from noise.
Moreover, the tall bars for the segments that are about 500-days long 
    point to the configurations
    that did not yield any changepoints. 
This is a satisfying observation indicating that
    there is some \emph{long-term stability} in the studied performance metrics,
    and therefore, specific types of performance experiments
    can be run repeatedly over long periods of time without 
    noticeable impact on the results caused by infrastructure changes and transient effects.
We further discuss several exemplary cases in the following section.     

\section{Investigating Change Patterns}
\label{investigating}

In this section we describe the most noticeable performance changes
    that we detect using CPD
    and investigate them by looking at the history of CloudLab system changes.
The latter comes in the form of kernel, OS, and compiler version
    changes and other attributes we record in our dataset,
    as well as the record of maintenance procedures
    (reconstructed based on administrators' recollections and emails sent to testbed users).
The cases where we can attribute the multi-benchmark changepoints 
    to such system changes provide \emph{validation},
    giving us the concrete context for the observed performance changes.
We present the summary of the instances we have investigated in Table~\ref{table:cpd-findings} and discuss them below.

\begin{table}[t]
    \centering
    \caption{Details of Validated CPD Results.}
    \resizebox{\columnwidth}{!}{
    \begin{tabular}{|l|l|l|l|}
        \hline
        \textbf{Hardware} & \makecell{\textbf{Change} \\ \textbf{Direction}} & \textbf{Time} & \textbf{Summary} \\
        \hline
        \texttt{xl170} & \makecell{CPU $\downarrow$, Mem $\uparrow$} & Nov 1, 2019 & \makecell{BIOS Updates;\\(see Fig.~\ref{fig:xl170})} \\ \hline
        \texttt{c6320}, other hw & \makecell{Mem $\downarrow$} & August 13, 2018 & \makecell{Upgrade from Ubuntu 16 to 18;\\(see Fig.~\ref{fig:c6320})} \\ \hline
        \texttt{d430}, other hw & \makecell{CPU $\uparrow$, Mem $\downarrow$} & July 25, 2019 & \makecell{Kernel upgrade from\\ \texttt{4.15.0-47} to \texttt{4.15.0-55}} \\ \hline
    \end{tabular}}
    \label{table:cpd-findings}
    \vspace*{-0.4cm}
\end{table}

\subsection{Major Changepoints}

\begin{figure*}[ht]
        \centering
        \begin{subfigure}[b]{0.65\textwidth}
            \centering
            \includegraphics[clip, trim=0.0cm 1.35cm 0.95cm 1.1cm,
            height=3.0cm, keepaspectratio]{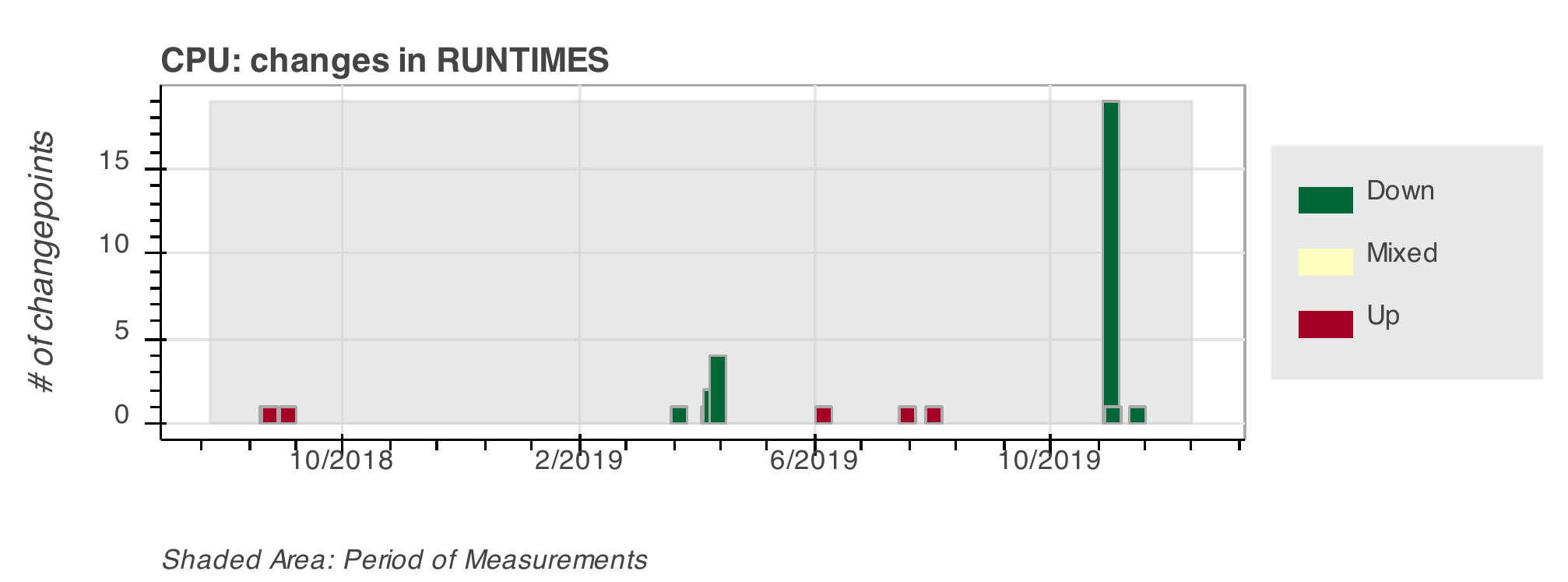}
            \caption[cpu-changepoints]%
            {{\small Changepoints in CPU traces.}}    
        \end{subfigure}
        \hfill
        \begin{subfigure}[b]{0.275\textwidth}  
            \centering 
            \includegraphics[clip, trim=0.0cm 0.45cm 0.0cm 1.1cm,
            height=3.0cm, keepaspectratio]{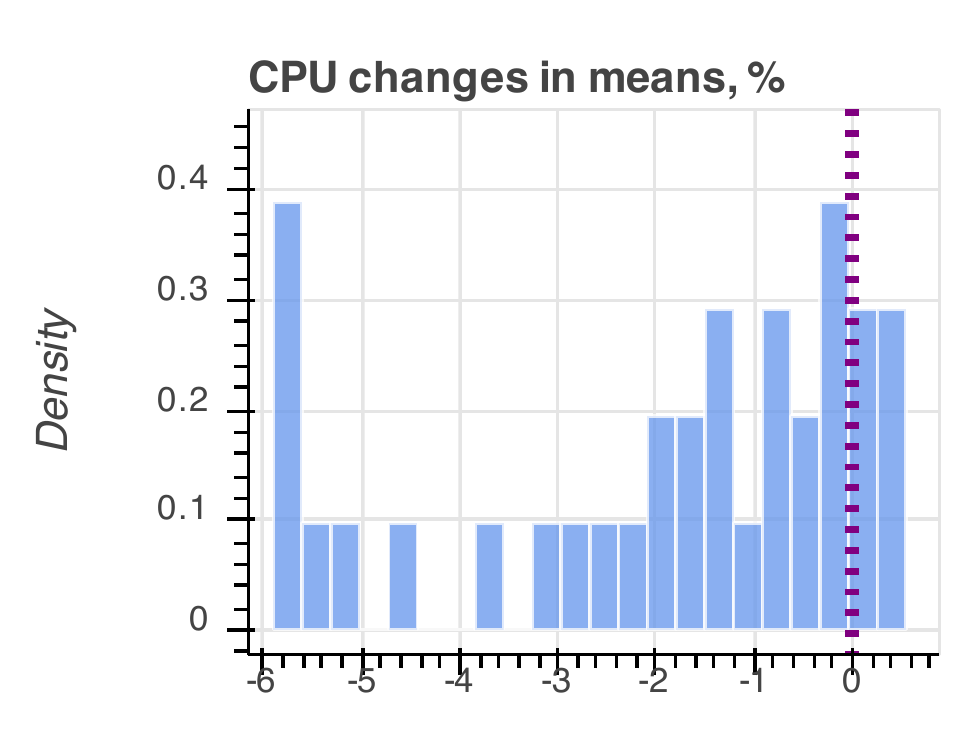}
            \caption[Magnitude of Change]%
            {{\small Histogram of CPU $c_n$ values.}}    
        \end{subfigure}
        \vskip\baselineskip
        \begin{subfigure}[b]{0.65\textwidth}   
            \centering 
            \includegraphics[clip, trim=0.0cm 1.35cm 0.95cm 1.1cm,
            height=3.0cm, keepaspectratio]{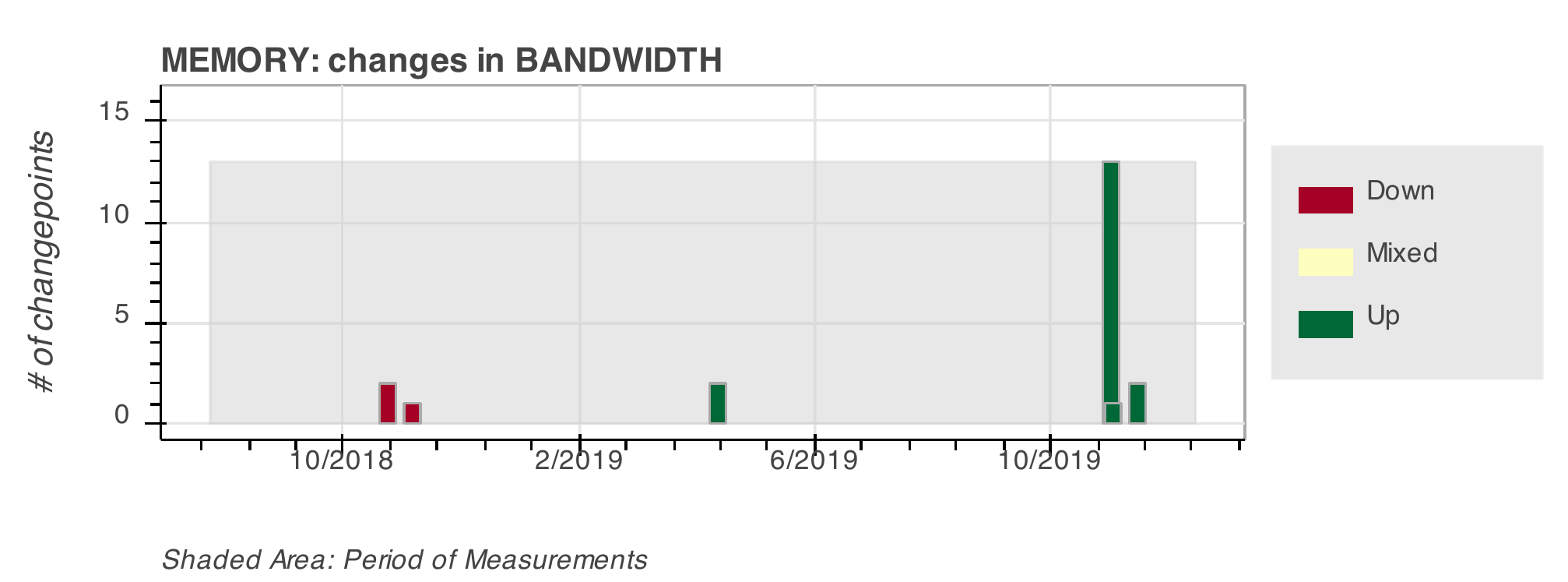}
            \caption[]%
            {{\small Changepoints in Memory traces.}}    
            \label{fig:mem-timeline}
        \end{subfigure}
        \hfill
        \quad
        \begin{subfigure}[b]{0.275\textwidth}   
            \centering 
            \includegraphics[clip, trim=0.0cm 0.45cm 0.0cm 1.1cm,
            height=3.0cm, keepaspectratio]{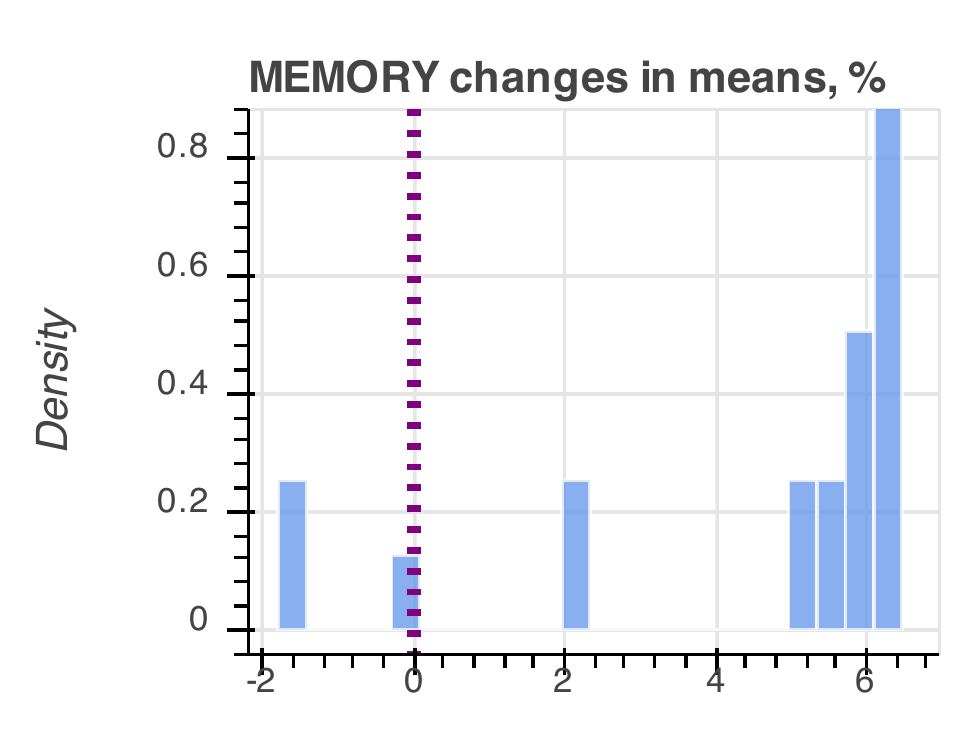}
            \caption[]%
            {{\small Histogram of Memory $c_n$ values.}}    
            \label{fig:mem-hist}
        \end{subfigure}
        \caption[ Text ]
        {\small Changepoint detection for \texttt{xl170} performance data. Shaded areas in (a) and (c) represent the period of benchmarking.} 
        \label{fig:xl170}
    \end{figure*}

\boldpar{xl170 BIOS Updates} 
In Figure~\ref{fig:xl170}, we illustrate how \texttt{xl170} performance traces show changes that occurred after November 1, 2019. 
The timeline plots (a,c) indicate that the changes can be considered \emph{positive}: CPU runtimes decreased, while memory bandwidth results increased (these changes can also be seen in Figure~\ref{fig:cpd}). Considering that the bars in these timelines represent individual
days, we confirm that the large spikes in the numbers of affected tests match the dates of updates to the BIOS conducted administrators by on these machines. We see that 
they are followed by several days with small numbers of related performance changes. 
This can be explained by a combination of how we collect our benchmarking results---we may get only a few machines tested on a day when the testbed and this particular hardware type are in high demand---and the fact that some of the tests might need more measurements
than the others to results in changepoints, depending on the magnitudes of changes. 
In this case, however, the detection is quite accurate;
    it points us precisely at the performance impact of system changes. 
    
The specific changes that caused this changepoint were tree BIOS settings according 
    to the HPE's low-latency tuning recommendations~\cite{hp-wp}.
By disabling patrol scrubbing (which scans memory to correct soft errors) and the early warning of DRAM errors (through the memory pre-failure notification setting), administrators reduced the amount of System Management Interrupts (SMIs) sent to the processor.  Administrators also reduced the rate at which the memory controller refreshes DRAM,
    from 2x to 1x.

As far the magnitudes of these hardware-specific changes are concerned,
    on average, CPU runtimes decreased by $3.1\%$ ($5.9\%$ maximum),
    and memory bandwidth increased by $6.0\%$ ($6.5\%$ maximum).
In Figure~\ref{fig:xl170} (b,d), we show the histograms for the full ranges of \texttt{xl170}'s $c_n$ values---not only for these BIOS-related
    changes but also for the changepoints that occurred at different 
    times---characterizing the entire period of our benchmarking for \texttt{xl170} machines.

\begin{figure*}[ht]
        \centering
        \begin{subfigure}[b]{0.475\textwidth}
            \centering
            \includegraphics[clip, trim=0.4cm 1.35cm 0.95cm 1.1cm,
            height=2.4cm, keepaspectratio]{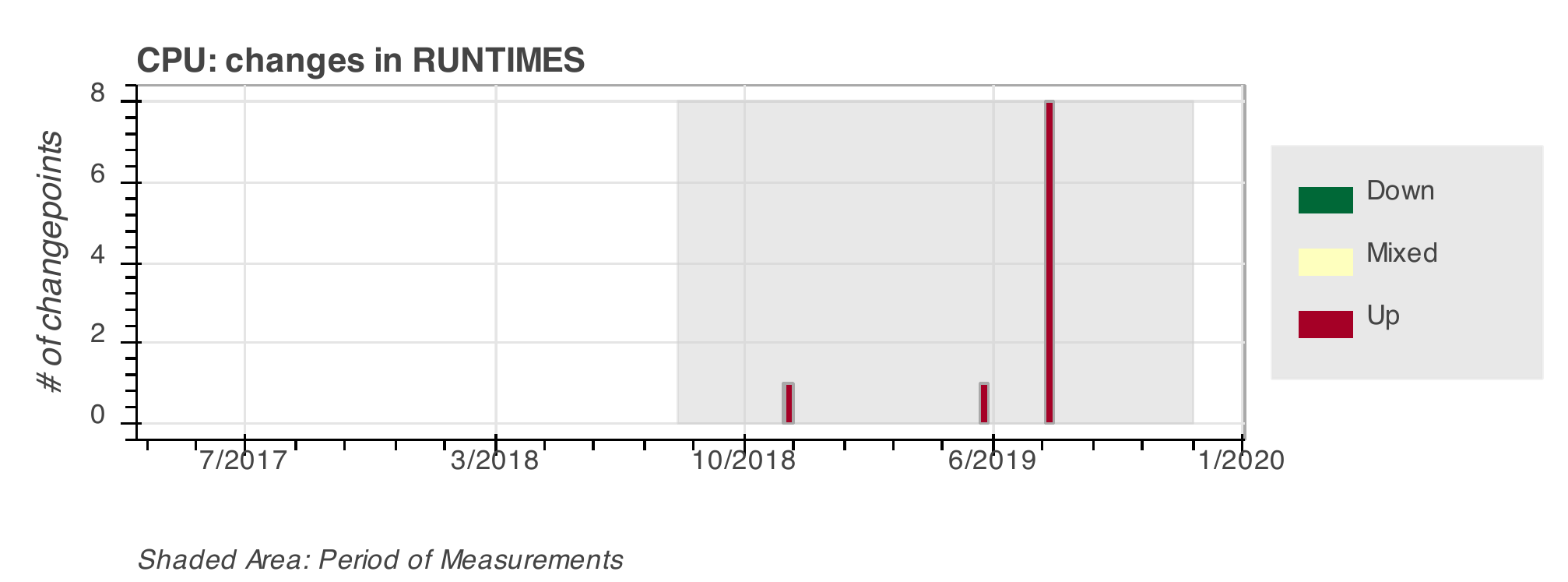}
            \caption[cpu-changepoints]%
            {{\small Changepoints in CPU traces.}}    
        \end{subfigure}
        \hfill
        \begin{subfigure}[b]{0.475\textwidth}  
            \centering 
            \includegraphics[clip, trim=0.4cm 1.35cm 0.95cm 1.1cm,
            height=2.4cm, keepaspectratio]{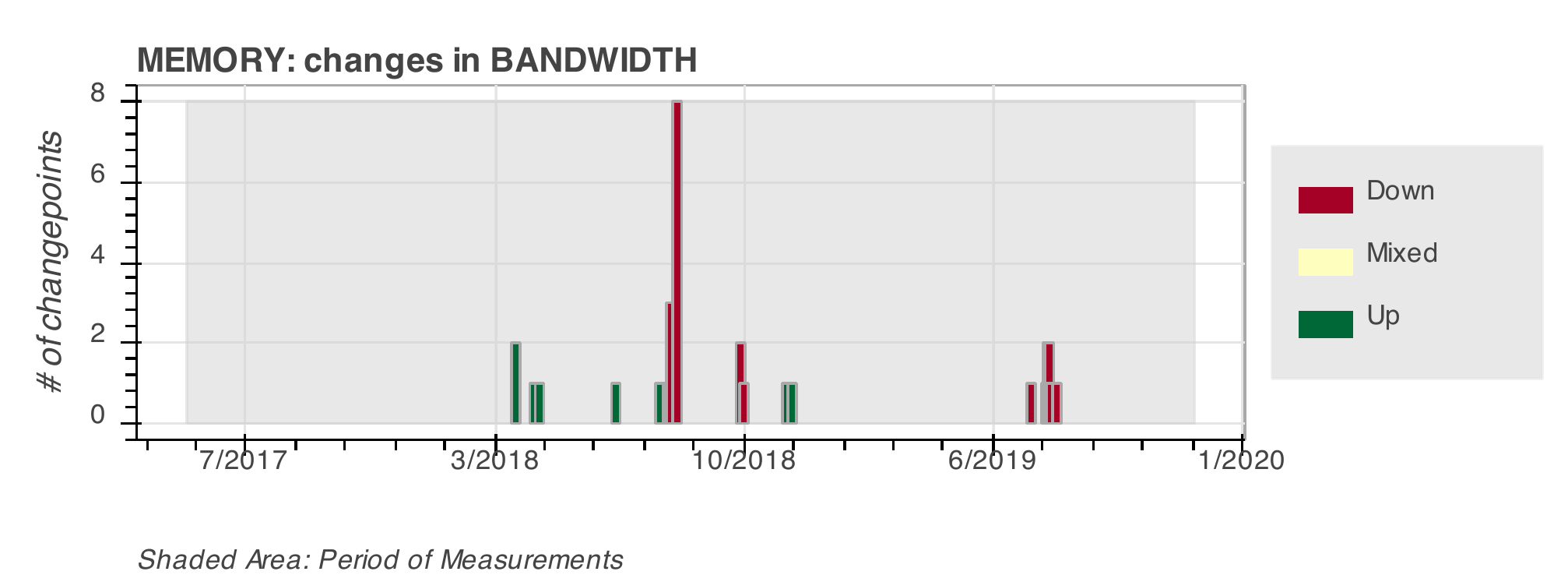}
            \caption[Magnitude of Change]%
            {{\small Changepoints in Memory traces.}}    
        \end{subfigure}
        \caption[ Text ]
            {\small Changepoint detection for \texttt{c6320} performance data. Shaded areas represent the period of benchmarking.} 
        \label{fig:c6320}
    \end{figure*}

\boldpar{OS Version Change} 
The performance effects of the testbed-wide switch from Ubuntu 16.04 as a default operating system image to Ubuntu 18.04 can be seen in the traces for most of CloudLab's hardware types,
    such as the \texttt{c6320} hardware type shown in Figure~\ref{fig:c6320}.
While our collection of CPU tests only began shortly after this transition,
    our memory results reveal the changepoints 
    that trace back to this OS upgrade.
These performance changes are predominantly \emph{negative}:
    they reflect the security updates
    many of which mitigate recent speculative execution exploits~\cite{kocher2019spectre} at the expense of performance.
\texttt{c6320}'s memory bandwidth results decreased by $2.7\%$ on average ($8.4\%$
maximum) following the OS switch, but not all hardware types experienced this degree of change. The \texttt{m510} hardware type, in contrast, showed smaller 
performance degradation, with the average of $0.8\%$ ($2.9\%$ maximum).
In the most stable case, \texttt{c220g1} machines 
    showed a single memory changepoint with bandwidth degradation of only $0.2\%$.
    
This helps illustrate the point that 
    we cannot project performance results from one hardware type to another.
This conclusion is typically drawn for high-level applications,
    yet here we see the evidence for it in the context of performance baselines defined by the OS evolution.

\boldpar{Kernel Version Changes} 
Similar to the OS upgrade, changes in the version of the deployed Linux kernel
    result in many performance changepoints.
One such update---from \texttt{4.15.0-47} to \texttt{4.15.0-55}---can
    be seen in the traces for multiple hardware types, 
    for both CPU and memory measurements.
A number of the corresponding \emph{negative} changepoints around July 25, 2019,
    which relate to the continuing mitigation of security exploits
    (we have confirmed that the changelogs for this series of kernel updates
    describe such improvements),
    can be seen in Figure~\ref{fig:c6320}.
There, CPU runtimes increased by $1.4\%$ on average ($2.2\%$
maximum), and memory bandwidth decreased by $1.7\%$ on average ($2.0\%$
maximum).
The largest relative change that we can attribute to this update
    is the $5.1\%$ memory bandwidth decrease 
    for read AVX instructions run on \texttt{c220g5} machines. 
It is also worth noting that the discussed kernel version update
    coincided with a compiler change, from GCC \texttt{7.3.0} to \texttt{7.4.0}.
To verify the root cause of these changepoints,
    we collected measurements in a series of tests that used a CloudLab machine 
    running the older kernel and the newer GCC.
By comparing these results with the pre- and post-changepoint distributions,
    we can claim without doubt that 
    the kernel is indeed responsible for the performance impacts, not GCC.
Our analysis of other kernel version changes, which took place
    before and after the aforementioned update,
    showed that they had more limited performance effects.
    
\subsection{Stable Measurements}

In addition to investigating changepoints, we 
    recognize long steady states present in our performance measurements
    and describe several instances below.

\boldpar{Disk Performance}
The most stationary measurements come from 
    the disks installed on \texttt{c6420} and \texttt{6220}.
Both use Seagate 1TB 7200-RPM 6G SATA HDDs (albeit different models).
Their performance traces showed isolated changepoints 
    for I/O tests with the default setting $iodepth=1$.
These are instances of impressive long-term performance stability,
    considering that other storage devices,
    such as Micron M500 120GB SATA3 flash
    and Toshiba XG3 series 256GB NVMe disks,
    showed 15 and 33 changepoints, respectively, on the same set of eight I/O tests.
This summary agrees with what we have found in our previous work~\cite{Maricq+:OSDI18}
    about the performance of these devices based on 
    empirical coefficients of variance:
    increased performance often comes at the expense of increased variability.
We also notice that I/O performance changepoints mostly do not coincide 
    with the rest of the studied changepoints,
    comparing to the CPU and memory changepoints that are in agreement in many cases,
    as discussed earlier.
    
\boldpar{Most Stable Configurations} 
CPU measurements on \texttt{m400} machines
    and memory measurements on \texttt{c6420} machines
    showed no changepoints.
Though, it does not mean that there is no variability or subtle fluctuations:
    we do find some performance changepoints when we increase $K$,
    but their number is still lower than the numbers of 
    changepoints detected for other hardware.
Based on all our observations, 
    these hardware types would be the best candidates---not showing the highest performance but instead delivering the highest stability---for long-running series of experiments with the emphasis on CPU and memory performance, among the hardware available on the CloudLab testbed. 
In such experiments, \emph{application performance regressions}
    can be studied in isolation from 
    hardware- and OS-caused transient performance effects,
    with greater rigor and depth.
    
\subsection{Discussion}

We anticipate that these findings can be helpful to the community of
    CloudLab users in several ways.
First, it is worth noting that some of the factors that caused changes
    (the OS and kernel updates) are actually under the users' control: while
    the user may opt to use the images and kernels that are the default at
    the time, they also have the option of using specific (potentially older)
    software stacks if performance consistency is a primary goal.
Second, as noted above, some hardware types show fewer changepoints than
    others, underscoring the fact that the choices the user makes of
    hardware can affect how many changes in performance they experience.
Third, it is notable that spinning disks provide the most consistent
    performance over time.
If this holds for larger set of models and usage patterns,
    this suggests that practical performance fingerprinting
    for performance explainability should put more emphasis on benchmarking of CPUs, 
    memory, and other types of I/O devices comparing to testing spinning disks.
    
While the analysis in this section necessarily concentrates on the 
    specific changes that we have observed in CloudLab,
    the overall lessons can generalize to other systems.
While changes are common in large, long-lived facilities, techniques
    such as CPD can help both administrators and users recognize these
    changes and track down their root causes.


\section{Tuning The Detection Process}
\label{tuning}

\begin{figure}[t]
    \centering
      \includegraphics[clip, width=0.35\textwidth]{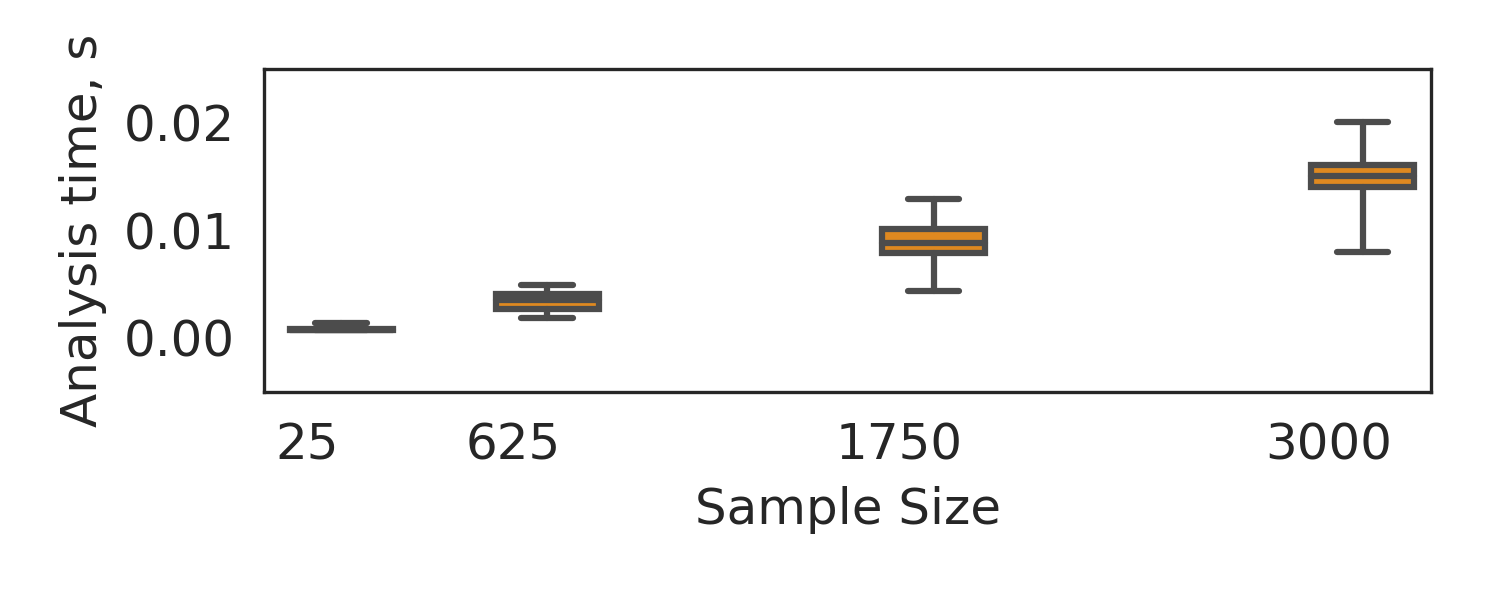}
    \caption{Time needed for single-timeseries CPD analysis.}
    \label{fig:single-cpd-runtime}
    \vspace*{-0.5cm}
\end{figure}

\begin{figure}
    \captionsetup[subfigure]{justification=centering}
    \centering
    \begin{subfigure}{0.5\textwidth}
      \centering
      \includegraphics[clip, width=1.0\textwidth]{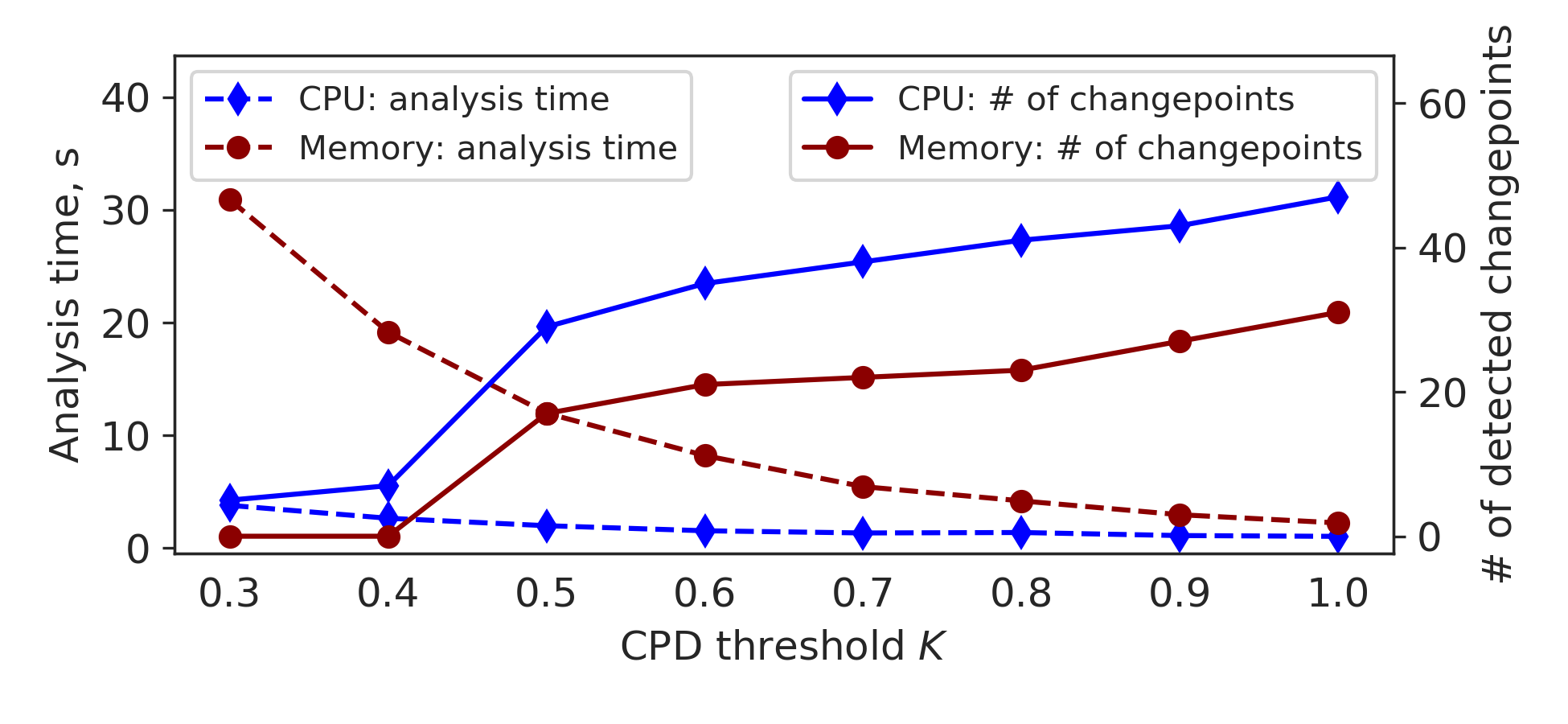}
      \caption{Properties of CPD for \texttt{x1l170} data.}
      \label{fig:tuning-xl170}
    \end{subfigure}%
    \vskip\baselineskip
    \vspace*{-0.35cm}
    \begin{subfigure}{0.5\textwidth}
      \centering
      \includegraphics[clip, width=1.0\textwidth]{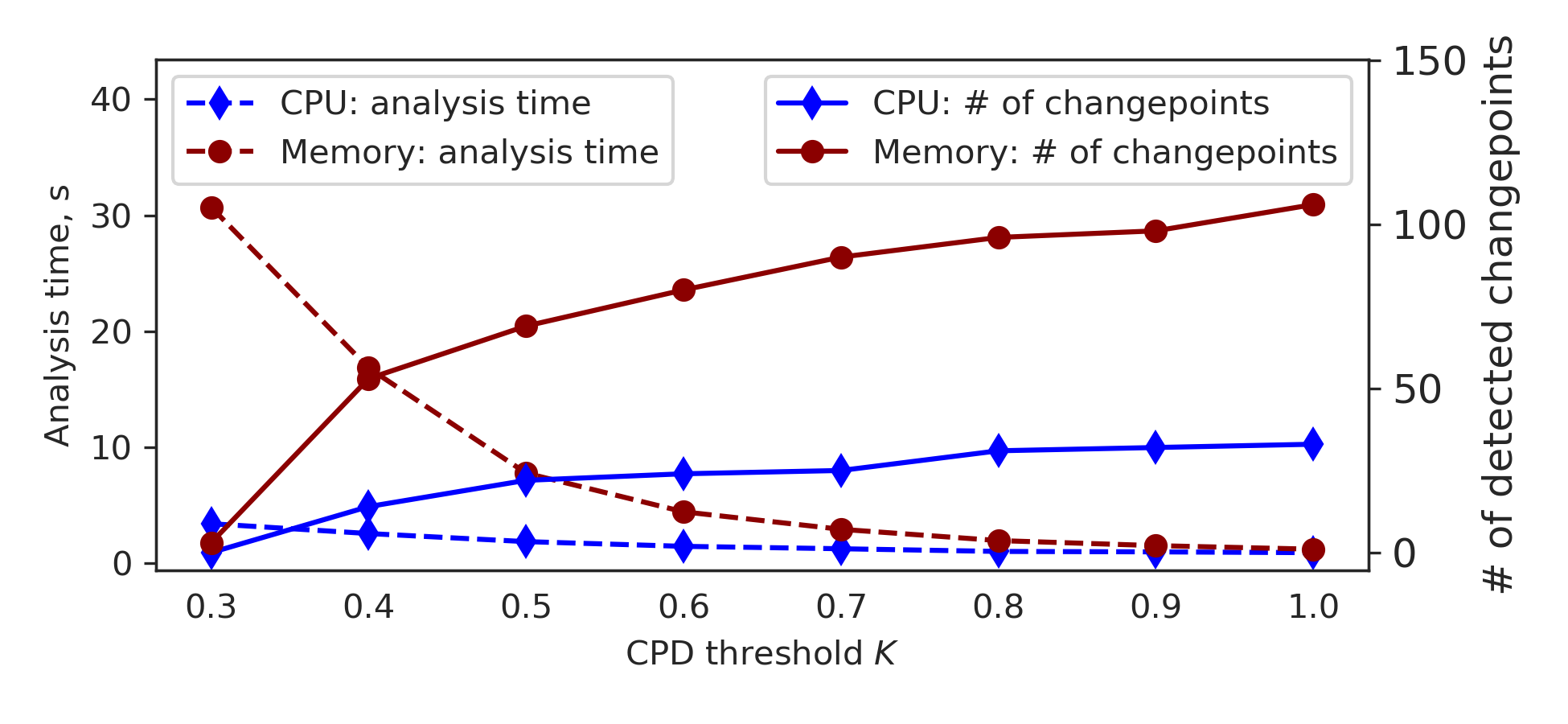}
      \caption{Properties of CPD for \texttt{c220g1} data.}
      \label{fig:tuning-c220g1}
    \end{subfigure}
    \caption{Results of varying the $K$ threshold.}
    \label{fig:tuning}
    \vspace*{-0.2cm}
\end{figure}

As part of our evaluation of CPD's computational requirements,
    we perform two series of experiments:
    CPD for single timeseries and 
    for batches of timeseries.
In the former, 
    we run the analysis for samples
    of measurements of the increasing size
    and measure per-invocation CPD analysis time (in seconds). 
We start with the 25-point sample shown in    
    Figure~\ref{fig:cpd}~(c) and proceed to larger 
    samples drawn from the data shown in Figure~\ref{fig:cpd}~(b). 
Each sample is randomly shuffled 
    (so we can study a variety of segmentations)
    and passed to the CPD code
    100 times.
We depict the analysis times we have recorded using
    the boxplot in Figure~\ref{fig:single-cpd-runtime}.
In the latter set of experiments, 
    which is more representative of the scenarios with multi-benchmark performance analysis for datacenters,
    we run CPD for \emph{all} CPU timeseries back-to-back,
    and then repeat it for \emph{all} memory data.
We study how the analysis times for these two batches
    vary as we tune the detection by changing the 
    value of $K$ threshold.
In Figure~\ref{fig:tuning}, we show the runtimes,
    as well as the numbers of identified changepoints,
    producing visualizations that allow us to
    reflect on the previously used value of $K$.
All these runtimes are collected and processed on a machine with two 6-core Xeon X5650 processors and 96 GB of memory.
Below we summarize the key insights 
    revealed by the performance results we have gathered. 

\boldpar{CPD is fast.}
It processes samples with over 3,000 points within
    tiny fractions of a second. 
The analysis time is linear with respect to the sample size.
Moreover, unlike typical Machine Learning tools that 
    require large datasets for training,
    CPD analyzes each sample independently from the rest of the data and can achieve good results 
    even on small, 25-point samples (as we demonstrated in Figure~\ref{fig:cpd}~(c)).
This makes CPD and the \texttt{robseg} implementation in particular suitable for fast, interactive analysis tools connected
to live databases or sources of streaming data.

\boldpar{There is a ``sweet spot'' in the  range of $K$ values.} 
Analysis time consistently decreases as we increase $K$, while the number of detected changepoints increases.
This matches what is highlighted in the study that introduced \texttt{robseg}~\cite{fearnhead2019changepoint},
where the authors compared two scenarios:
``no change'' ($K=0$) and ``many changes'' ($K=n/100$, where $n$ is the sample size). Our results complement their brief summary
by demonstrating how CPD behaves over an entire range of $K$ values 
that may be considered in practice. Thus, thresholds around $K=0.6$ appear
to be good choices for CPU and memory performance analysis,
as confirmed by the trade-off curves shown in Figure~\ref{fig:tuning}.
Such values allow finding most of the changepoints detected with higher values of $K$
without performance drawbacks of the analysis with lower $K$.
It is also worth noting that if we are indeed interested in making the detection more sensitive by increasing $K$,
we would be able to do that without performance penalties.
Then, we would need to select $K$ based on
    the desired characteristics of changepoints 
    and steady states, 
    using the arguments from Section~\ref{cpd}.

\section{Open Artifacts}
\label{artifacts}

\boldpar{CONFIRM} is an interactive analysis service running at
    \texttt{\url{https://confirm.fyi/}}.
    We developed it to assess the levels of performance and variability 
    present in our dataset using scatter plots, per-machine confidence intervals, 
    converging overall confidence intervals, among other analysis techniques.
    It analyzes all performance results we have collected on CloudLab.
    It is worth noting that we first encountered performance
    changepoints while exploring specific configurations one-by-one using CONFIRM's scatter plots, which are similar to the ones shown in Figure~\ref{fig:cpd} (without CPD segmentations). 
    
\boldpar{Change Over Time} is a complementary dashboard we have developed to 
    examine the results of CPD.
It is available at \texttt{\url{https://confirm.fyi/change/}}.
    It runs alongside CONFIRM, accesses the same database, and produces visualizations like the ones shown in Figures~\ref{fig:xl170} and~\ref{fig:c6320}.
    The dashboard summarizes the numbers of detected changepoints and their 
    distributions for CPU, memory, and disk measurements, 
    as well as depicts the temporal relationships between them.
    The dashboard has a slider allowing to experiment with 
    the CPD threshold $K$ and go 
    from a few (only large) to the larger numbers (including smaller) of performance changepoints.

\boldpar{All data and the developed analysis code} 
can be found at: 
\href{https://gitlab.flux.utah.edu/emulab/cloudlab-ccgrid20}{\texttt{https://gitlab.flux.utah.edu/emulab/\\cloudlab-ccgrid20}}. This repository provides access to
the raw data, the collection of changepoints produced by our analysis, and 
the Google Colab notebooks that can be easily cloned and run.

\section{Impact of Changepoint Analysis}
\label{impact}

With the approach we have laid out and the findings we have presented,
    we hope to demonstrate the value of CPD for comprehensive performance evaluation and operation of large computing facilities.
Not only can such analysis improve facility studies in 
    cloud computing, HPC, datacenter optimization, etc., 
    but it can also contribute to a wide range of practical
    evolution-over-time studies focused on 
    performance of individual components and aspects of modern cyberinfrastructure:
    new devices, network and storage systems, compilers and computing frameworks, QoS policies, 
    resource sharing, among others.
Better understanding of all these systems requires
    measuring key performance characteristics 
    across a variety of operational regimes rather than focusing on a single or handful of selected states.
    
A good analysis protocol would prescribe 
    checking stationarity across the sampled states
    or time intervals and using elements of 
    CPD where the stationarity does not persist.
The same logic applies to computing applications
    used repeatedly in production settings.
We support this protocol and have shared what we have learned from it 
    when we discussed the stationarity assessment in our previous work~\cite{Maricq+:OSDI18}
    and described the changepoint investigation in the current study.
    
In a different capacity,
    we can also envision the use of CPD
    in the systems \emph{reproducibility} studies, 
    which reuse published data and code artifacts
    for validation and further analysis.
CPD can be a part of the toolchains involved in such work, complementing     other temporal analysis techniques.

The confluence of ideas about establishing performance baselines and studying reproducibility leads us to think about creating an open \emph{archive} for performance variability data.
There is plenty of precedent for archives of performance data, from the Top500~\cite{top500} and Green500~\cite{green500} efforts, to the data released by SPEC~\cite{spec}. What we propose is an archive specifically targeting performance stationarity and variability. This archive would include fine-grained multi-benchmark datasets, similar to the dataset we describe in this paper. It would provide access to complete performance records for repeated benchmark runs on the same system, as well as runs across groups of theoretically-identical systems. This would allow system owners and users better understand what to expect in their environments and bootstrap statistical calculations and adaptive techniques needed for robust benchmarking of their hardware.
An organized and documented collection of data like this
is likely to empower a plethora of new studies
on large-scale performance evaluations, analysis of correlations in performance measurements, outlier detection, prediction of results in untested configurations, minimization of benchmarking time, among many other
avenues within contemporary performance analysis of computer systems. 
In this context, CPD techniques running on the archive's data
would strengthen all analyses that are able to consider individual stationary segments and their performance characteristics rather than relying on the coarse estimates obtained for entire time series (sometimes with nonstationarities). 
We hope that our dataset from this study, in addition to the recently published measurements of network performance in several clouds~\cite{uta-nsdi20-zenodo} (also showing many changepoints), 
will help establish such archive and stimulate more work on novel and practical CPD.

\section{Conclusion and Future Work}
\label{conclusion}

In the current study, we apply changepoint detection to a large dataset of measurements we have collected on the CloudLab testbed,
which includes records of CPU, memory, and disk performance. 
We present our analysis of the detected changepoints---their distributions
in terms of magnitudes and inter-changepoint intervals---how they relate 
to the large recorded system changes, and also reveal which configurations we have found to be the most stable based on the lack of changepoints.
These results, coupled with the presented performance experiments and their outcomes, have convinced us in the viability and usefulness of applying robust CPD in studying large performance datasets. 
We expect to see more work in computer systems research using CPD techniques 
in the future and have shared several ideas about the types of developments
that might facilitate this adaptation.

As part of our future work, we plan to include the results of network bandwidth and latency tests
    collected on CloudLab.
It will allow us to compare CloudLab's networks
    with the networks deployed in public clouds~\cite{uta2020NSDI}
    from the variability and evolution-over-time perspectives.
On the detection side, we will experiment with 
    other changepoint detection approaches and tools, including 
    online Bayesian detection~\cite{adams2007bayesian}
    and the BreakoutDetection package~\cite{breakout},
    looking for the best choices and capabilities
    for datacenter performance analysis.
We will work on a comprehensive comparison study for the 
    available methods and consider evaluating them on the CloudLab's
    performance dataset.

\section*{Acknowledgments}

This material in this paper is based upon work supported by the National Science Foundation, Grant Number 1743363.

\bibliographystyle{IEEEtran}
\bibliography{refs}

\end{document}